\documentclass[conference]{IEEEtran}
\IEEEoverridecommandlockouts
\usepackage{cite}
\usepackage{amsmath,amssymb,amsfonts}
\usepackage{bm}
\usepackage{algorithmic}
\usepackage{graphicx}
\usepackage{textcomp}
\usepackage{xcolor}
\def\BibTeX{{\rm B\kern-.05em{\sc i\kern-.025em b}\kern-.08em
    T\kern-.1667em\lower.7ex\hbox{E}\kern-.125emX}}
\makeatletter

\def\ps@IEEEtitlepagestyle{%
  \def\@oddfoot{\mycopyrightnotice}%
  \def\@evenfoot{}%
}
\def\mycopyrightnotice{%
  {\footnotesize\hfill}
  \gdef\mycopyrightnotice{}
}
\begin{document}
\newcommand{\grad}{\bm\nabla}
\newcommand{\minmod}{\mathsf{mm}}
\newcommand{\maxp}{\mbox{$\max^+$}}
\newcommand{\sgn}{\mbox{sgn}}
\newtheorem{theorem}{Theorem}

\def\citepunct{,}

\title{Stabilised Inverse Flowline Evolution for Anisotropic Image 
Sharpening\\
\thanks{
This project has received funding from the European Research Council (ERC) 
under the European  Union’s Horizon 2020 research and innovation programme 
(grant agreement No. 741215, ERC Advanced Grant INCOVID).
}
}
\author{\IEEEauthorblockN{Kristina Schaefer}
\IEEEauthorblockA{\textit{Mathematical Image Analysis Group} \\
\textit{Saarland University}\\
Saarbr\"ucken, Germany \\
schaefer@mia.uni-saarland.de }
\and
\IEEEauthorblockN{Joachim Weickert}
\IEEEauthorblockA{\textit{Mathematical Image Analysis Group} \\
\textit{Saarland University}\\
Saarbr\"ucken, Germany \\
weickert@mia.uni-saarland.de }
}
\maketitle
\begin{abstract}
The central limit theorem suggests Gaussian convolution as a generic 
blur model for images. Since Gaussian convolution is equivalent to 
homogeneous diffusion filtering, one way to deblur such images is to 
diffuse them backwards in time. However, backward diffusion is highly 
ill-posed. Thus, it requires stabilisation in the model as well as 
highly sophisticated numerical algorithms. Moreover, sharpening is 
often only desired across image edges but not along them, since it 
may cause very irregular contours. This creates the need to model 
a stabilised anisotropic backward evolution and to devise an
appropriate numerical algorithm for this ill-posed process.

We address both challenges. First we introduce stabilised 
inverse flowline evolution (SIFE) as an anisotropic image sharpening flow.
Outside extrema, its partial differential equation (PDE) is backward
parabolic in gradient direction. Interestingly, it is sufficient to 
stabilise it in extrema by imposing a zero flow there.  
We show that morphological derivatives -- which are not common in the
numerics of PDEs -- are ideal for the numerical approximation
of SIFE: They effortlessly approximate directional derivatives
in gradient direction. Our scheme adapts one-sided morphological
derivatives to the underlying image structure. It allows to progress
in subpixel accuracy and enables us to prove stability properties. Our 
experiments show that SIFE allows nonflat steady states and outperforms 
other sharpening flows.
\end{abstract}

\begin{IEEEkeywords}
mathematical morphology, morphological derivatives, 
backward parabolic PDEs, image enhancement

\end{IEEEkeywords}

\section{Introduction}
Image processing with evolution equations based on partial differential 
equations (PDEs) offers many successful tools with numerous applications, 
including image denoising \cite{Ii62}, enhancement \cite{PM90,We98}, and 
variational restoration \cite{AK06}. 
These methods are often inspired by physics. For instance, diffusion-based 
image processing takes its inspiration from the physical diffusion 
process. Its evolution equations are forward parabolic PDEs and have 
smoothing properties \cite{We98}. Conversely, diffusion backward in 
time has a sharpening effect. This can be useful for tasks such as 
contrast enhancement or deblurring, when no explicit blur kernel is 
known or blind deconvolution methods such as \cite{BCLG19} are to
be avoided. First image processing papers in this
direction date back to 1955~\cite{KJ55}. Unfortunately, such processes 
are ill-posed: In the same way as forward diffusion attenuates high 
frequencies exponentially, backward diffusion amplifies them exponentially. 
Without additional stabilisation, the image range explodes within a
short time. It took decades to come up with first successful solutions for
this very difficult problem. A minimalistic regularisation to
tame the explosive behaviour is pursued in stabilised inverse linear 
diffusion (SILD) \cite{OR91}. It uses backward linear diffusion, but 
forbids over- and undershoots at extrema. Its sophisticated discretisation 
requires one-sided finite differences with minmod switches.
Unfortunately, it sharpens not only across image edges, but also
along them, which creates unpleasant irregular contours. 

\medskip
\textbf{Contributions.} The goal of our paper is build upon these ideas 
and improve them  by introducing an anisotropic version of SILD.
We advocate {\em stabilised inverse flowline evolution (SIFE)} as a novel 
PDE for image sharpening.  
Outside extrema, this filter is based on an anisotropic backward parabolic 
PDE in gradient direction. This anisotropy together with the ill-posedness 
of backward parabolic evolutions makes it very challenging to devise an 
adequate numerical algorithm. We show that by replacing finite differences 
with morphological derivatives, we can derive a scheme that is simple and 
elegant. We prove that it is stable w.r.t. a maximum--minimum 
principle. Thus, it cannot produce over- and undershoots.
Our experiments show that SIFE allows nonflat steady states and 
performs better than SILD and shock filtering.   

\medskip
\textbf{Related Work.}
In terms of image sharpening flows, our SIFE filter most
closely resembles its predecessor SILD~\cite{OR91}.
Both are backward parabolic, ill-posed, and benefit from
a stabilisation at extrema. SILD, however, is linear and isotropic,
while SIFE is nonlinear and anisotropic. Image analysis with PDEs
that contain anisotropic backward parabolic terms have a long
tradition \cite{Ga65,GSZ02a,WW21}, but they are typically stabilised
by forward parabolic terms. Even with these stabilisations it remains 
nontrivial to find provably stable numerics. A backward parabolic 
1-D model with range stabilisation and a simple explicit
scheme has been introduced recently \cite{BCWW20}. It is not 
applicable to our anisotropic 2-D problem.
Shock filters 
\cite{KB75,OR90} are hyperbolic alternatives for morphological 
image sharpening. Their wave-like propagation is less challenging
w.r.t.~ill-posedness.
To our knowledge, a
fully backward parabolic 2-D anisotropic model has not been studied
so far.

\medskip
PDE-based processes are typically discretised with finite differences. 
This also includes shock filters and SILD. Due to its anisotropic and 
nonlinear nature, discretising SIFE with finite differences is 
numerically challenging. A specific feature of our numerical algorithm
is its usage of morphological derivative 
approximations~\cite{Be90,LHS87,RSB93,VYB89}. So far, morphological 
derivatives are mainly used for feature detection. Apart from an 
image prior for denoising~\cite{Na15}, we are not aware of any 
applications within implementations of PDE-based filters.

\medskip
In recent years many deep learning approaches for image sharpening have
been proposed \cite{ZRLL22}. They involve millions 
of hidden parameters and usually
perform well without offering formal guarantees or deeper insights into
their inner workings. Our goal is to come up with a transparent model with
a single parameter, and to understand how to design provably stable 
numerical algorithms for it.
Thus, comparing both complementary philosophies w.r.t.~a single criterion
such as performance or stability guarantees would not do justice to any 
of them. Both are needed, but under different requirements.

\medskip
\textbf{Paper Structure.} 
Section \ref{sec:morph} reviews the relevant background from mathematical 
morphology, in particular w.r.t. morphological derivatives. In Section 
\ref{sec:sife}, we introduce SIFE and derive its numerical approximation 
with morphological derivatives. We compare the performance of SIFE with 
other image enhancing flows in Section~\ref{sec:experiments}. 
Section \ref{sec:conclusion} offers conclusions and an outlook to future 
work.


\section{Review of Mathematical Morphology}
\label{sec:morph}

Since morphological derivatives are essential for approximating the
PDE-based SIFE filter, we first have to review necessary prerequisites
from mathematical morphology.


\subsection{Dilation and Erosion}
Mathematical morphology is a classical and very successful area of image 
processing~\cite{So04}. Its building blocks are dilation and erosion. 
The \emph{dilation} $\delta_B[f]$ of a continuous greyscale image $f$ 
with a \emph{structuring element} $B$ -- a set that typically characterises 
the notion of a neighbourhood -- replaces 
each value of $f$ by its supremum within the structuring element. The 
erosion $\varepsilon_B[f]$ uses the infimum instead. Formally, the operations 
are defined as
\begin{align}
\label{eq:dil}
\delta_B[f](\bm x) &\;=\; \sup \, \lbrace f(\bm x - \bm b)\;|\; \bm b \in 
 B\rbrace,\\ \label{eq:ero}
\varepsilon_B[f](\bm x) &\;=\; \inf \, \lbrace f(\bm x + \bm b)\;|\; \bm b 
 \in B\rbrace. 
\end{align}
In this paper, we only use ball-shaped structuring elements, i.e.~symmetric 
intervals for 1-D signals, and disks for images. This guarantees rotation
invariance and aligns morphological derivatives in gradient direction. In 
the following, we write $\delta_r$ and $\varepsilon_r$ for dilations and 
erosions with a ball of radius $r$.

\medskip
It has been shown that dilation / erosion $u(\bm x, t)$ of $f$ with a 
ball of radius $t$ solves the PDE
\begin{equation}
\label{eq:morphology-pde}
\partial_t u \;=\; \pm |\grad u|
\end{equation}
with initial condition $u(\bm x, 0) = f(\bm x)$; see e.g.~\cite{BM92}. 
Dilation uses the plus sign and erosion the minus sign. By $\bm{\nabla}$ we
denote the spatial nabla operator, and $|\,.\,|$ is the Euclidean norm.


\subsection{Morphological Derivatives}

Dilations and erosions allow morphological approximations
of image derivatives. To this end, one computes differences 
between the original image, its dilations, and its erosions. For 
instance, the {\em internal} and {\em external 
gradient}~\cite{LHS87,RSB93,Ma05} with a ball-shaped structuring 
element of radius $r$ are defined as
\begin{align}
\mbox{internal gradient:} & 
   &g_r^-[f](\bm x)\; &= \; \frac{f(\bm x) - \varepsilon_{r}[f](\bm x)}{r},
   \\[1mm]
\mbox{external gradient:} & 
   &g_r^+[f](\bm x)\; &= \;\frac{\delta_{r}[f](\bm x) - f(\bm x)}{r}.
\end{align}
Averaging both operators yields the {\em Beucher 
gradient}~\cite{Be90,RSB93}.

\medskip
Let us consider $f$ at some location $\bm x$ with nonvanishing gradient.
It has been shown \cite{Ma05} that for $r\to 0$, the preceding morphological 
gradients converge to the derivative $f_\eta$ in {\em flowline direction}  
$\bm \eta = \frac{\grad f}{|\grad f|}$: 
\begin{align}
\lim_{r\to 0}\: g_r^-[f](\bm x) \; &= \; f_\eta,\\
\lim_{r\to 0}\: g_r^+[f](\bm x) \; &= \; f_\eta.
\end{align}

\medskip
Internal, external, and Beucher gradient show a structural resemblance 
to forward, backward, and central finite difference approximations
of first order derivatives. For finite differences, we can construct 
approximations of higher order derivatives by computing the difference 
of approximations of lower order derivatives. This strategy is also 
applicable to morphological derivative approximations. 
For instance, the difference between internal and external 
gradient yields a morphological approximation of the 
second directional derivative $f_{\eta\eta}$ \cite{VYB89}.


\subsection{Rouy--Tourin Scheme for PDE-based Morphology}
For digital images, we need discrete analogues for dilation and
erosion.
One can directly discretise the set-theoretic description
\eqref{eq:dil}--\eqref{eq:ero} by replacing the supremum with the maximum, 
and the infimum with the minimum. However, discretising the PDE 
\eqref{eq:morphology-pde} offers subpixel accuracy and an improved 
rotation invariance for small structuring elements. Therefore, we 
pursue the PDE-based approach in our paper.

\medskip
For this purpose, we use the Rouy--Tourin upwind scheme~\cite{RT92}. 
For a derivation of this scheme from \eqref{eq:dil}--\eqref{eq:ero} 
with purely geometric arguments, we refer to van den Boomgaard~\cite{Bo99}.
Let $u_{i,j}^k$ denote the grey value of pixel $(i,j)$ at time level 
$k$, and assume that the
pixel size is given by $h$, and the time step size by $\tau$. Then
the Rouy--Tourin scheme approximates a dilation step with a disk of
radius $\tau$ by
\begin{equation}
\label{eq:dil-numerics}
\resizebox{0.89\linewidth}{!}{
$\begin{aligned}
\frac{u^{k+1}_{i,j} - u^{k}_{i,j}}{\tau}
  = \Big(
      &\max\Big\lbrace 0,\, \frac{u^{k}_{i+1,j} - u^{k}_{i,j}}{h},\,
                  \frac{u^{k}_{i-1,j} - u^{k}_{i,j}}{h}
          \Big\rbrace^2 \\
      +&\max\Big\lbrace 0,\, \frac{u^{k}_{i,j+1} - u^{k}_{i,j}}{h},\, 
                    \frac{u^{k}_{i,j-1} - u^{k}_{i,j}}{h}
          \Big\rbrace^2\Big)^{\frac{1}{2}}.
\end{aligned}$
}
\end{equation}
Initialising this iterative scheme with the original image 
$u^0_{i,j} = f_{i,j}$ and performing $k$ iterations approximates a
dilation with a disk of radius $k\tau$. In 2D, one can show that 
this scheme satisfies the maximum--minimum principle
\begin{equation}
 \min_{m,n} f_{n,m} \;\leq\; u_{i,j}^k \;\leq\; \max_{m,n} f_{n,m} 
 \qquad \mbox{for all $i$, $j$}
\end{equation}
provided that $\tau \leq \frac{h}{\sqrt{2}}$. 
In 1D, the time step step size limit is $\tau \leq h$. 
Analogously, erosion is approximated by
\begin{equation}
\label{eq:ero-numerics}
\resizebox{0.89\linewidth}{!}{
$\begin{aligned}
\frac{u^{k+1}_{i,j} - u^{k}_{i,j}}{\tau}
  = - \Big(
      &\max\Big\lbrace  0,\, \frac{u^{k}_{i,j} - u^{k}_{i+1,j}}{h},\,
                   \frac{u^{k}_{i,j} - u^{k}_{i-1,j}}{h}
          \Big\rbrace^2\\
      +&\max\Big\lbrace 0,\, \frac{u^{k}_{i,j} - u^{k}_{i,j+1}}{h},\,
                    \frac{u^{k}_{i,j} - u^{k}_{i,j-1}}{h}
          \Big\rbrace^2\Big)^{\frac{1}{2}}.
\end{aligned}$
}
\end{equation}

\medskip
After this discussion of morphological concepts we are now in a position
to study our novel PDE for image sharpening.  


\section{Stabilised Inverse Flowline Evolution}
\label{sec:sife}
In this section we introduce the stabilised inverse flowline evolution 
(SIFE) as a flow for image sharpening. It is motivated by the
stabilised inverse linear diffusion (SILD) of Osher and Rudin \cite{OR91}.
However, while SILD is a linear and isotropic process that is approximated 
by finite differences, SIFE is nonlinear and anisotropic, and it benefits 
from morphological derivatives. First we derive the PDE for SIFE. 
Afterwards we propose a stable numerical algorithm with morphological
derivatives in 1D, and we extend it to arbitrary dimensions.


\subsection{PDE Formulation of SIFE}

Filters based on forward parabolic PDEs such as diffusion methods 
are widely used in image processing~\cite{Ii62,We98}. For
a continuous image $f(\bm x)$, isotropic linear diffusion \cite{Ii62}
creates a family $\,\{u(\bm x,t)\:|\:t \ge 0\}\,$ of smoothed versions 
by solving  
\begin{align}
\partial_t u = \Delta u
\end{align}
with initial condition $u(\bm x, 0) = f(\bm x)$ and reflecting boundary 
conditions. Here $\Delta$ denotes the spatial Laplacian.
The smoothness increases with the diffusion time $t$. 

\medskip
While forward diffusion has smoothing properties, backward diffusion 
sharpens images by reversing smoothing. In contrast to forward parabolic 
processes, backward parabolic evolutions are typically ill-posed and 
need a stabilisation to avoid instabilities such as the violation of 
a maximum--minimum principle. 

\medskip
Osher and Rudin proposed such a stabilisation. Their 
{\em stabilised inverse linear diffusion (SILD)}~\cite{OR91}
regularises backward diffusion by fixing values at extrema: 
\begin{align}
\partial_t u \;=\; -\sgn(|\grad u|)\,\Delta u\,.
\end{align}
The sign factor implements the stabilisation, since $|\grad u|$ 
vanishes in extrema. This enforces a maximum--minimum principle.
SILD has one obvious drawback: For image sharpening, one is usually 
only interested in sharpening across edges, but not along them. 
The latter can cause unpleasant irregular contours. Being an 
isotropic process, however, SILD does not allow such a 
direction-specific filtering.  

\medskip
As a remedy, we propose an anisotropic modification of SILD that
acts only perpendicular to edges, i.e.~in flowline direction
$\bm{\eta} = \frac{\bm{\nabla} u}{|\bm{\nabla} u|}$. This leads to
the equation
\begin{equation}
\label{eq:sife-cont}
\partial_t u \;=\; -\sgn(|\grad u|) \, \partial_{\eta\eta} u
\end{equation}
which we call \emph{stabilised inverse flowline evolution (SIFE)}.
Note that unlike SILD, SIFE is nonlinear, since $\bm{\eta}$ depends
on the evolving local image structure $u(\bm{x},t)$.


\subsection{Numerical Challenges of SIFE}

While it is easy to write down the SIFE PDE, finding stable algorithms is 
very challenging for two reasons:
\begin{enumerate}
\item Equation \eqref{eq:sife-cont} is backward parabolic and hence ill-posed.
      Thus, naive discretisations immediately create instabilities.
      As a remedy, we will adapt the sophisticated minmod scheme for 
      SILD~\cite{OR91} to our purpose.
\item The anisotropy poses additional problems. A classical way to handle
      it would be to use 
      \begin{equation}
        u_{\eta\eta} \;=\;
       \frac{u_x^2 \, u_{xx} \:+\:
       2\, u_x \:  u_y \:  u_{xy}
       \:+\; u_y^2 \,  u_{yy}}
       {u_x^2 \:+\: u_y^2}
      \end{equation}
      and approximate all derivatives on the right-hand side with finite
      differences. Unfortunately, it is completely unclear how this can 
      be combined with the minmod concept of SILD. Here, morphological 
      derivatives
      can rescue us, since they effortlessly approximate derivatives in
      flowline direction in any dimension.
\end{enumerate} 
We now proceed with these ideas by first translating the SILD numerics
based on finite differences into a stable SIFE scheme with morphological
derivatives. This is done in the 1-D setting, where we also prove
stability results. Afterwards we generalise the morphological scheme to 
higher dimensions.


\subsection{Morphological SIFE Numerics in 1D}

In 1D, the {\em minmod finite difference scheme}
for SILD \cite{OR91} is given by   
\begin{equation}
\resizebox{0.89\linewidth}{!}{
$\begin{aligned}
 \label{eq:sild-discrete}
 \frac{u^{k+1}_{i} - u_i^k}{\tau} \;=\;
  -\frac{1}{h} \, &\Biggl( \: \minmod 
     \, \biggl( \frac{u_{i+2}^k - u_{i+1}^k}{h}, \,
            \frac{u_{i+1}^k - u_{i}^k}{h}, \,
            \frac{u_{i}^k - u_{i-1}^k}{h} \biggr)\\
    \quad &-\minmod  \, \biggl( \frac{u^k_{i+1} - u^k_{i}}{h}, \,
            \frac{u^k_{i} - u_{i-1}^k}{h}, \,
            \frac{u^k_{i-1} - u^k_{i-2}}{h} \biggr) 
   \Biggr)
\end{aligned}$
}
\end{equation}
with initial condition $\bm{u}^0 = \bm{f}$. To implement reflecting
boundary conditions, the discrete image is extended by a mirrored 
boundary layer of size two pixels.
The minmod function $\minmod$ returns the 
argument with the smallest absolute value if all of them have equal sign, 
and $0$ otherwise. In extrema, forward and backward differences have
different signs, such that the minmod function yields zero.
Hence, grey values of extrema are frozen in time. This is essential 
for the stability of the discrete evolution.

\medskip
To transfer this stabilisation concept to SIFE in arbitrary dimensions, 
we replace finite differences by morphological derivatives. This 
yields simple approximations of the directional derivative 
$\partial_{\eta\eta} u$. We start with the 1-D case.

\medskip
For a 1-D signal, we propose the following morphological scheme for
SIFE:

\begin{equation}
\resizebox{0.89\linewidth}{!}{
$\begin{aligned}
 \label{eq:sife-discrete}
 \frac{u^{k+1}_{i} - u_i^k}{\tau} = 
  -\frac{1}{r}\Biggl(&\min\,
    \biggl( \frac{\delta_{2r}[u]_{i}^k - \delta_{r}[u]_{i}^k}{r}, \,
            \frac{\delta_{r}[u]_{i}^k - u_{i}^k}{r}, \,
            \frac{u_{i}^k - \varepsilon_{r}[u]_{i}^k}{r} \biggr) \\
    -&\min\,\biggl( \frac{\delta_{r}[u]_{i}^k - u_{i}^k}{r},\,
            \frac{u_{i}^k - \varepsilon_{r}[u]_{i}^k}{r}, \,
            \frac{\varepsilon_{r}[u]_{i}^k - \varepsilon_{2r}[u]_{i}^k}{r} 
            \biggr)\Biggr)
\end{aligned}$
}
\end{equation}
with reflecting boundaries and the initial signal $\bm u^0 = \bm f$. The 
operations $\delta_{r}$ and $\varepsilon_r$ denote 1-D dilation and erosion
with a symmetric 1-D structuring element. Since 
morphological first order derivatives approximate $u_\eta = |\grad u|$, they
are always nonnegative. Thus, the minmod becomes the minimum. 
In (\ref{eq:sife-discrete}), we need dilations and erosions with radii
$r$ and $2r$. We compute them with the Rouy--Tourin scheme 
(\ref{eq:dil-numerics}), (\ref{eq:ero-numerics}). Its stability condition 
restricts us to radii of $r\leq h$ in 1D. However, if one aims at accurate 
derivative approximations, small radii should be used. Thus, we recommend
$r:=\frac{h}{2}$ and use this throughout our experiments. We perform one 
iteration for $\varepsilon_{r}$ and $\delta_{r}$, and two for 
$\varepsilon_{2r}$ and $\delta_{2r}$.

\medskip
It is not hard to see that outside extrema the SIFE scheme approximates 
its underlying PDE with first order consistency in space and time. In
extrema it reproduces the continuous model exactly by preventing any 
changes. This also implies that the continuous SIFE model and its
algorithm do not alter binary images, since they only consist of extrema
and already offer maximal sharpness.


\subsection{Stability of the 1-D SIFE Scheme}

For a disk radius $r=h$, SIFE is equivalent to SILD in 1D. Thus, it is not 
surprising that SIFE offers the same stability properties that have been 
claimed for SILD in 1D. However, no proof for SILD is given in \cite{OR91}, 
and establishing stability results also for $r<h$ -- which is relevant for 
us -- is technically more demanding. Below is our stability theorem.
Since a detailed proof is very lengthy, we only sketch the main steps.
 
\medskip
\begin{theorem}[Stability of SIFE]
Let $\bm{f}$ be a discrete 1-D signal and $r \le h$. Consider the SIFE 
scheme \eqref{eq:sife-discrete}  with one Rouy--Tourin iteration 
for $\delta_r$ and $\varepsilon_r$, and two for $\delta_{2r}$ and 
$\varepsilon_{2r}$. If $\tau \leq r^2$, the following properties 
hold for all $k \ge 0$:
\vspace{1mm}
\begin{enumerate}
\item[(a)] \textbf{Preservation of Monotonicity} 
  \begin{align}
     &f_{i-1} \leq f_{i} \quad \forall i 
      \quad \Rightarrow \quad u^k_{i-1} \leq u^k_{i} \quad \forall i,\\ 
     &f_{i-1} \geq f_{i} \quad \forall i 
      \quad \Rightarrow \quad u^k_{i-1} \geq u^k_{i} \quad \forall i.
  \end{align}
\item[(b)] \textbf{Maximum--minimum Principle} 
  \begin{align}
    \min_{j} f_{j} \;\leq\; u_{i}^k \;\leq\; \max_{j} f_j \quad \forall i.
  \end{align}
\end{enumerate}
\end{theorem}

\medskip
\emph{Sketch of the Proof.} 
First we prove monotonicity preservation. Let $\bm{u}^k$ be locally
concave and increasing (other cases can be treated in a 
similar way). Thus, we have $u^k_{i+1}-u^k_{i} \leq u^k_{i}-u^k_{i-1}$ 
and $u_{i-1}^k \leq u_{i}^k$. Then the Rouy--Tourin 
dilations and erosions in \eqref{eq:sife-discrete} yield 
\begin{align*}
\delta_{2r}[u]_{i}^k &\,=\, u_i^k + \frac{2r}{h}(u_{i+1}^k - u_i^k) 
    + \frac{r^2}{h^2}(u_{i+2}^k - 2u_{i+1}^k + u_i^k), \\[1mm]
\delta_{r}[u]_{i}^k &\,=\, u_i^k + \frac{r}{h}(u_{i+1}^k - u_i^k),\\[1mm]
\varepsilon_{r}[u]_{i}^k &\,=\, u_i^k - \frac{r}{h}(u_{i}^k - u_{i-1}^k),\\
\varepsilon_{2r}[u]_{i}^k &\,=\, u_i^k - \frac{2r}{h}(u_{i}^k - u_{i-1}^k) 
    + \frac{r^2}{h^2}(u_{i-2}^k - 2u_{i-1}^k + u_i^k).
\end{align*} 
This gives the differences 
\begin{align*}
\delta_{2r}[u]_{i}^k - \delta_{r}[u]_{i}^k 
  &\;\leq\; \delta_{r}[u]_{i}^k - u_i^k 
  \;\leq\; u_i^k - \varepsilon_{r}[u]_{i}^k,\\
\delta_{r}[u]_{i}^k - u_i^k
  &\;\leq\; u_i^k - \varepsilon_{r}[u]_{i}^k 
  \;\leq\; \varepsilon_{r}[u]_{i}^k - \varepsilon_{2r}[u]_{i}^k.
\end{align*}
With that, SIFE results in 
\begin{align*}
u^{k+1}_{i} \:=\: u_i^k 
    - \tau\,\frac{\delta_{2r}[u]_{i}^k - 2 \delta_{r}[u]_{i}^k +u_i^k}{r^2}\,.
\end{align*}
Since we also assume concavity of $\bm u^k$ in $i-1$, we get 
\begin{align*}
u^{k+1}_{i-1} \:=\: u_{i-1}^k 
    - \tau\,\frac{\delta_{2r}[u]_{i-1}^k - 2 \delta_{r}[u]_{i-1}^k 
    + u_{i-1}^k}{r^2}\,.
\end{align*}
Subtracting both equations gives
\begin{equation}
\label{eq:proof-difference-concave}
\begin{aligned}
u^{k+1}_{i} - u^{k+1}_{i-1} 
  &\;=\; (u_i^k - u^{k}_{i-1} ) \Big(1- \frac{\tau}{r^2}\Big)\\
  &\;-\; \frac{\tau}{r^2}(\delta_{2r}[u]_{i}^k - \delta_{2r}[u]_{i-1}^k)\\[1mm]
  &\;+\; \frac{2\tau}{r^2}(\delta_{r}[u]_{i}^k - \delta_{r}[u]_{i-1}^k).
\end{aligned}
\end{equation}
In a next step one shows that
\begin{align}
\label{eq:dil-inequality}
\delta_{2r}[u]_{i}^k - \delta_{2r}[u]_{i-1}^k 
   \;\leq\; \delta_{r}[u]_{i}^k - \delta_{r}[u]_{i-1}^k
\end{align}
holds by plugging in the results for dilation and erosion, and 
exploiting the local concavity of $\bm{u}^k$ together with $r \leq h$.
Plugging \eqref{eq:dil-inequality} in \eqref{eq:proof-difference-concave}, 
we obtain 
\begin{align*}
u^{k+1}_{i} - u^{k+1}_{i-1} &\;\geq\;
  (u_i^k - u^{k}_{i-1} ) \Big(1- \frac{\tau}{r^2}\Big) \\
  &\;\quad\; + \frac{\tau}{r^2}\,
     (\delta_{r}[u]_{i}^k - \delta_{r}[u]_{i-1}^k)\\ 
  &\;\geq\; 0\,,
\end{align*}
since $\bm{u}^k$ and $\delta_{r}[\bm{u}^k]$ are increasing and 
$\tau \leq r^2$. Therefore, monotonicity is preserved in this case.

\smallskip
If $\bm u^k$ has a maximum in $i$, then SIFE results in
\begin{align*}
u^{k+1}_{i} &\:=\: u_i^k, \\
u^{k+1}_{i-1} &\:=\: u_{i-1}^k + 
  \tau\,\frac{\delta_{r}[u]_{i-1}^k +u_{i-1}^k}{r^2}.
\end{align*}
The difference between the two equations yields
\begin{align*}
u^{k+1}_{i} - u^{k+1}_{i-1} \:=\; 
  (u_i^k - u^{k}_{i-1} ) - \tau\frac{\delta_{r}[u]_{i-1}^k +u_{i-1}^k}{r^2}
 \:\geq\: 0
\end{align*}
for $\tau \leq r^2$, since $\delta_{r}[u]_{i-1}^k \leq u_i^k$ if $\bm u^k$ 
has a maximum in $i$. Thus, the monotonicity is preserved in the 
increasing case. The decreasing case in analogous. 

\smallskip
Monotonicity preservation implies the maximum--minimum principle:
Extrema are not changed by the evolution, and monotone regions remain 
monotone. Therefore, values cannot increase past maxima or decrease below 
minima. 
~\hfill$\square$

\subsection{Morphological SIFE Numerics in Higher Dimensions}

The extension of the SIFE scheme to higher dimensions is straightforward:
One simply needs to use a higher dimensional, ball-shaped structuring 
element, e.g. a disk in case of an image. The stability limit of the 
Rouy-Tourin scheme restricts us to radii $\,r \leq h/\sqrt{2}\,$ in 2D.
Also in 2D we recommend $r=h/2$, a single Rouy--Tourin
iteration for $\varepsilon_r$ and $\delta_r$, and two iterations
for $\varepsilon_{2r}$ and $\delta_{2r}$. This offers a good compromise
between subpixel accuracy and efficiency. 

\medskip
Interestingly, in our experiments the 2-D SIFE scheme satisfies a 
maximum--minimum principle for time step sizes up to $r^2$. 
Thus, its stability limit does not deteriorate when going from 1D to 2D. 
This is uncommon for explicit discretisations of PDE-based methods. SILD, for 
instance, must reduce its stability limit from $\tau \leq \frac{h^2}{2}$ 
to $\tau \leq \frac{h^2}{4}$ \cite{OR91}. 
However, in contrast to SILD, the extreme anisotropy of SIFE turns it into
a pure 1-D sharpening flow: It only acts in flowline direction. 
If this is the reason why its 1-D stability limit transfers to higher 
dimensions without further restrictions, also stability reasonings 
favour SIFE over SILD.

\section{Experiments}
\label{sec:experiments}

Let us now evaluate the performance of SIFE by comparing it to two PDE
evolutions that have been designed for the same purpose: SILD and shock 
filtering. A widely used shock filter \cite{KB75,OR90} is governed 
by the hyperbolic evolution
\begin{equation}
\partial_t u = -\sgn(\Delta u) |\grad u|
\end{equation}
with initial condition $u(\bm x, 0) = f(\bm x)$ and reflecting boundary 
conditions. The sign of $\Delta u$ signals whether we are in the
influence zone of a maximum $(\Delta u < 0)$ or a minimum $(\Delta u > 0)$.  
This determines if we perform dilation or erosion. Shocks develop
at the interface between adjacent influence zones.
 
\medskip
For our experiments, we discretise the shock filter with the Rouy--Tourin
scheme. Our grid size $h$ is assumed to be $1$. For the morphological 
derivatives in SIFE we choose a disk of radius $r=0.5$ as structuring 
element. This subpixel accuracy is achieved with a single Rouy--Tourin 
step. For SIFE and SILD we use the time step size $\tau = 0.2$, and for 
shock filtering $\tau = 0.5$.

\begin{figure}[tb]
\centering
\begin{footnotesize}
\begin{tabular}{cc}
original & shock filter \\[0.5mm]
\includegraphics[scale=0.435]{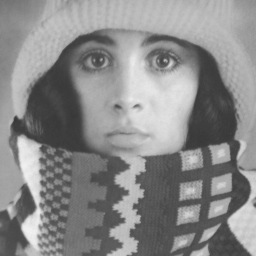} & 
\includegraphics[scale=0.435]{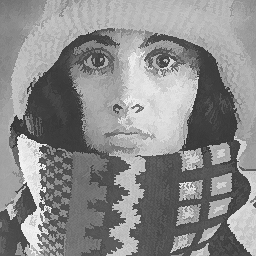}\\[3.5mm] 
\includegraphics[scale=0.435]{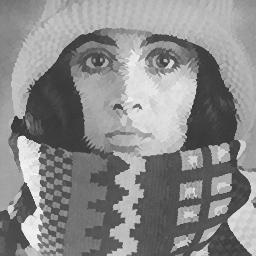} &
\includegraphics[scale=0.435]{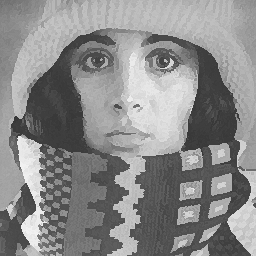}\\
SILD & SIFE
\end{tabular}
\end{footnotesize} 

\caption{Comparison of the steady states of shock filtering, SILD, 
  and SIFE for the $256\times 256$ greyscale image {\em trui}.}
\label{fig:exp-comparison}
\end{figure}

\begin{figure}
\centering
\begin{footnotesize}
\begin{tabular}{ccc}
\includegraphics[width=0.28\linewidth]{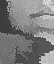} &
\includegraphics[width=0.28\linewidth]{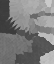} &
\includegraphics[width=0.28\linewidth]{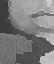}\\
shock filter & SILD & SIFE  
\end{tabular}
\end{footnotesize} 
\caption{Zoom into a $54\times 64$ window of 
 Figure \ref{fig:exp-comparison}.}
\label{fig:zoom}
\end{figure}

\medskip
In Figure \ref{fig:exp-comparison}, we compare the steady states 
($t \to \infty$) of SIFE, SILD, and shock filtering. 
While all approaches produce nonflat steady states with piecewise constant
segmentations, they differ substantially in detail. This is best visible 
in the zooms in Figure~\ref{fig:zoom}. 

\medskip
Interestingly, the backward parabolic 
SILD and SIFE processes produce more regular edges than the hyperbolic shock 
filter. We conjecture that this is caused by the fact that SILD 
and SIFE employ a $5 \times 5$ stencil to approximate a second order 
differential operator with first order accuracy. The minmod or minimum 
switches aim at small absolute derivative approximations, which is 
needed for regularising the ill-posed backward parabolic evolution. 
The hyperbolic shock filter, however, can be discretised adequately 
within a $3 \times 3$ neighbourhood. It is steered by a conventional 
$5$-point stencil that approximates the Laplacian with second order 
accuracy.

\medskip
From a practical perspective, the more regular boundary structure
of SILD and SIFE makes these backward parabolic sharpenings preferable
over the hyperbolic shock filter. This demonstrates that backward
parabolic evolutions deserve more attention than the more widely
used hyperbolic processes.

\medskip
It should be noted that SIFE has the most pleasant edge structure 
due to the absense of a backward parabolic term {\em along} edges: 
Its anisotropy allows to sharpen only {\em across} edges (i.e.~in gradient 
direction $\bm{\eta}$). In contrast, the isotropic sharpening of SILD 
causes irregular edges, since backward parabolic evolution along
edges is not prevented. This is also in agreement with Haralick's
observations in the context of edge detection~\cite{Ha84}, where he
argues that $u_{\eta\eta}$ should be favoured over $\Delta u$. 
In conclusion, Figures \ref{fig:exp-comparison}--\ref{fig:zoom} show 
that the SIFE model and its algorithmic implementation do exactly 
what they have been designed for.

\begin{figure}[tbp]
\centering
\begin{footnotesize} 
\begin{tabular}{cc}
\includegraphics[scale=0.22]{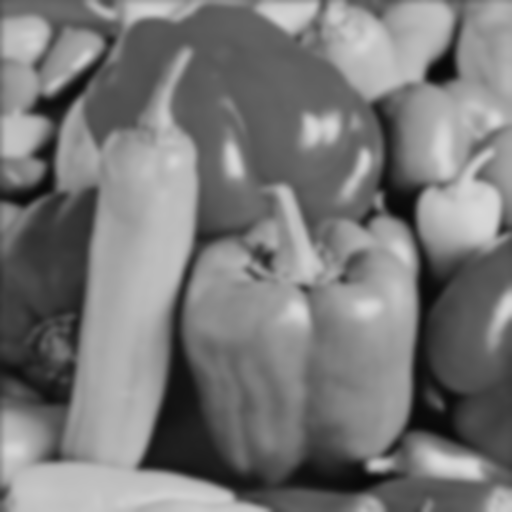} & 
\includegraphics[scale=0.22]{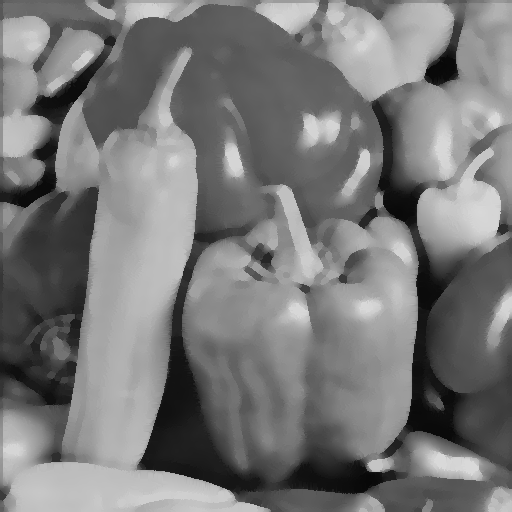}\\
 blurred image  & 
 sharpened image, $t=10$
\end{tabular}
\end{footnotesize} 
\caption{Image sharpening with SIFE. The original $512 \times 512$ 
 {\em peppers} image was degraded with Gaussian blur with $\sigma=3$.}
\label{fig:exp-sharp}
\end{figure}

\medskip
If we aim at image sharpening rather than a piecewise constant 
segmentation-like result, we should stop the evolution at a finite 
time. This is illustrated in Figure \ref{fig:exp-sharp} where SIFE 
nicely sharpens the Gaussian-blurred test image {\em pepper} without 
introducing staircasing artefacts. The runtime for processing 
this $512 \times 512$ image with 50 iterations on a Computer with 
Intel\textsuperscript{\textcopyright} Core\texttrademark \, i9-11900K 
CPU @ 3.50 GHz is 0.3 seconds.
 

\section{Conclusions}
\label{sec:conclusion}

We have proposed stabilised inverse flowline evolution (SIFE) as a novel 
image sharpening flow. It is goverend by an anisotropic
backward parabolic PDE in gradient direction that is stabilised at
extrema. Thus, its sharpening direction varies from location to
location, which makes the design of numerical algorithms challenging.
For its discretisation, morphological derivatives provide simple 
and elegant approximations of the directional derivative in 
gradient direction. Our experiments show that this filter can outperform 
other PDE-based processes for image sharpening such as stabilised inverse
linear diffusion and shock filtering.  

\medskip
In a more general setting, we have seen how one can handle
ill-posed nonlinear, and anisotropic backward parabolic PDEs in
a numerically adequate way. Thus, if such processes appear interesting 
from a modelling perspective, there is no reason to shy away from them. 
Moreover, our contributions may go far beyond concrete 
sharpening applications: It is very common to express PDE-based methods 
in their gauge coordinates $\bm{\eta} \parallel 
\bm{\nabla}u$ and $\bm{\xi} \perp \bm{\nabla}u$, e.g.~in the context
of level set methods and active contour models. Any progress on 
appropriate discretisations for differential operators of type 
$\,a\,\partial_{\eta\eta} + b\,\partial_{\xi\xi}$, where $a$ and $b$ may 
have any sign, is immediately applicable to all these problems. Our 
ongoing research is exploring extensions of SIFE towards this general 
formulation and other applications.

\medskip
To the best of our knowledge, our work is the first to explore the 
potential of morphological derivatives for numerical analysis: It 
shows that morphological derivatives are ideal tools to approximate
PDEs with derivatives in gradient direction. Moreover, it is
natural to combine them with advanced numerical ideas such as 
adaptive one-sided approximations. This may also pave the path for 
applications beyond image analysis, e.g.~in computational fluid 
mechanics where upwinding, minmod schemes, and WENO approximations 
are popular.

\medskip
{\bf Acknowledgement.} We thank Tobias Alt for useful comments on a
draft version of this paper.


\bibliographystyle{IEEEtran}
\bibliography{IEEEabrv,lit}
\end{document}